\documentclass[prb,amsmath,aps,showpacs,twocolumn,superscriptaddress]{revtex4}
\usepackage{graphicx,xspace}
\usepackage{float}
\usepackage{amsmath}
\usepackage{amssymb}

\input{epsf}

\usepackage{dcolumn}
\usepackage{bm}

\begin{document}

\title{Controlled nucleation of topological defects in the stripe domain patterns of  Lateral multilayers with Perpendicular Magnetic Anisotropy:
competition between magnetostatic, exchange and misfit
interactions}

\author{A. Hierro-Rodriguez}
\altaffiliation[Present address: ]{IN-IFIMUP, Depto. Fisica e
Astronomia, Faculdade de Ci\^{e}ncias, Universidade do Porto, Rua
Campo Alegre 687, 4169 - 007 Porto, Portugal} \affiliation{Depto.
F{\'i}sica, Universidad de Oviedo, 33007 Oviedo, Spain}
\affiliation{CINN, (CSIC-Universidad de Oviedo-Principado de
Asturias), Llanera, Spain}

\author{M. V\'elez}
\affiliation{Depto. F{\'i}sica, Universidad de Oviedo, 33007
Oviedo, Spain}

\author{R. Morales}
\affiliation{Depto. Qu{\'i}mica-F{\'i}sica \& BCMaterials,
Universidad del Pa{\'i}s Vasco-UPV/EHU, 48940 Leioa, Spain}
\affiliation{IKERBASQUE, Basque Foundation for Science, 48011
Bilbao, Spain}

\author{N. Soriano}
\affiliation{Depto. Qu{\'i}mica-F{\'i}sica, Universidad del
Pa{\'i}s Vasco-UPV/EHU, 48940 Leioa, Spain}

\author{G. Rodr{\'i}guez-Rodr{\'i}guez}
\affiliation{Depto. F{\'i}sica, Universidad de Oviedo, 33007
Oviedo, Spain}  \affiliation{CINN, (CSIC-Universidad de
Oviedo-Principado de Asturias), Llanera, Spain}

\author{L. M. \'Alvarez-Prado}
\affiliation{Depto. F{\'i}sica, Universidad de Oviedo, 33007
Oviedo, Spain}  \affiliation{CINN, (CSIC-Universidad de
Oviedo-Principado de Asturias), Llanera, Spain}
\author{J. I. Mart{\'i}n}
\email{jmartin@uniovi.es}\affiliation{Depto. F{\'i}sica,
Universidad de Oviedo, 33007 Oviedo, Spain} \affiliation{CINN,
(CSIC-Universidad de Oviedo-Principado de Asturias), Llanera,
Spain}
\author{J. M. Alameda}
\affiliation{Depto. F{\'i}sica, Universidad de Oviedo, 33007
Oviedo, Spain} \affiliation{CINN, (CSIC-Universidad de
Oviedo-Principado de Asturias), Llanera, Spain}

\begin{abstract}
Magnetic lateral multilayers have been fabricated on weak
perpendicular magnetic anisotropy amorphous Nd-Co films in order
to perform a systematic study on the conditions for controlled
nucleation of topological defects within their magnetic stripe
domain pattern. A lateral thickness modulation of period $w$ is
defined on the nanostructured samples that, in turn, induces a
lateral modulation of both magnetic stripe domain periods
$\lambda$ and average in-plane magnetization component
$M_{inplane}$. Depending on lateral multilayer period and in-plane
applied field, thin and thick regions switch independently during
in-plane magnetization reversal and domain walls are created
within the in-plane magnetization configuration coupled to
variable angle grain boundaries and disclinations within the
magnetic stripe domain patterns. This process is mainly driven by
the competition between rotatable anisotropy (that couples the
magnetic stripe pattern to in-plane magnetization) and in-plane
shape anisotropy induced by the periodic thickness modulation.
However, as the structural period $w$ becomes comparable to
magnetic stripe period $\lambda$, the nucleation of topological
defects at the interfaces between thin and thick regions is
hindered by a size effect and stripe domains in the different
thickness regions become strongly coupled.
\end{abstract}
%\pacs{}
\pacs{75.60.Jk, 75.70.Kw, 75.75.-c}

\maketitle

\section{INTRODUCTION}
Magnetic films with perpendicular magnetic anisotropy (PMA)
display a peculiar domain structure consisting of small regions
with up and down magnetization that can be arranged either in
regular stripe patterns or adopt many different beautiful
"labyrinthine" configurations along a hysteresis
loop.\cite{minor,loop} The actual domain pattern in a given PMA
film can be very complex depending on material parameters, sample
geometry and magnetic and thermal
history,\cite{molho,hehn,stripes3} but a simple description of
disordered stripe patterns can be achieved if the concept of
topological defects within the 2D periodic stripe magnetic
structure is used.\cite{stripes1,stripes2} Defects such as
dislocations,\cite{field} disclinations,\cite{stripes2} grain
boundaries\cite{huang} or even skyrmions\cite{skyrmion} have been
observed. These topological defects play an important role in
magnetization reversal processes and magnetization dynamics of PMA
materials\cite{venus,asciutto,defeo} and, also, in the physics of
phase transitions in 2D modulated phases.\cite{stripes3,saratz}
However, the experimental study of these topological defects in
PMA materials has been hindered by the problems to control their
nucleation in extended samples since they usually occur within
very disordered labyrinthine configurations.

 More recently, the idea of
topological defects within the magnetization configuration has
also been introduced  in order to describe domain walls in
magnetic nanostructures with in-plane
magnetization.\cite{walls,gabriel,walls2,walls3} Fractional
vortices near sample edges allow to understand many different
situations such as holes within a continuous magnetic
layer,\cite{gabriel} vortices in nanodots\cite{walls3} or domain
wall propagation in magnetic nanowires.\cite{walls2,parkin} In
this in-plane magnetization configuration, the restricted
nanostructure geometry allows for a good control of topological
defect nucleation and propagation
processes.\cite{walls,gabriel,walls2,walls3,parkin}

An ideal system to combine these two concepts of topological
defects can be found in weak PMA materials in which stripe domains
coexist with a significant in-plane magnetization component. When
perpendicular magnetic anisotropy $K_N$ becomes smaller than
magnetostatic energy ($E_{demag} = 2\pi M_s^2$ with $M_s$ the
saturation magnetization), weak stripe domains are nucleated in
the system\cite{huber} above a critical thickness. In this case,
the equilibrium domain configuration consists of a small
out-of-plane oscillation of the magnetization of amplitude $\Delta
M_{out}$ around an average in-plane magnetization $M_{inplane}$
that is aligned with the stripe direction due to the Bloch
character of the domain walls in between up and down domains (see
sketch in Fig.1(a)).\cite{huber,stripes_amplitude,clarke} From a
macroscopic point of view, coupling between in-plane and
out-of-plane magnetization components gives rise to an in-plane
pseudo-uniaxial anisotropy term called rotatable
anisotropy\cite{rotatable,luis} since in-plane magnetization
rotations imply a global reconfiguration of the whole stripe
pattern.

\begin{figure}[ht]
 \begin{center}
\includegraphics[angle=0,width=1\linewidth]{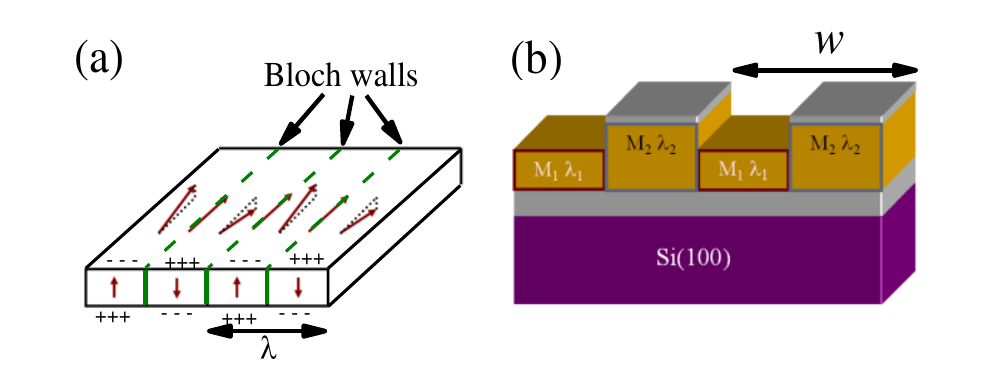}%ahierro_prb_fig1.eps}
 \end{center}
 \caption{(color online)(a) Sketch of magnetization configuration in the stripe domains of a weak PMA film. (b) Sketch of a PMA magnetic lateral
  multilayer: a thickness modulation of lateral period $w$ creates periodic changes in
  the average in-plane magnetization component $M_{i}$ and
  magnetic stripe domain period $\lambda_i$. } \label{Fabricacion}
 \end{figure}

Recently, experiments in  weak PMA magnetic lateral
multilayers\cite{aurelio} have shown an intrinsic coupling between
topological defects occurring within  the in-plane and
out-of-plane magnetization configurations (1/2 vortices and 1/2
disclinations) that could be of use to control the nucleation of
topological defects in the magnetic stripe domain pattern.
Briefly, a magnetic lateral multilayer (MLM) is a continuous
magnetic film with a lateral modulation of some relevant magnetic
property e.g. $M_S$,\cite{McCord,lateral} exchange
bias\cite{theis,theis2} or anisotropy.\cite{ingleses,jafaar} PMA
lateral multilayers (PMA-MLM) can be fabricated by introducing a
periodic thickness modulation\cite{aurelio} through a
nanostructuration process since the equilibrium stripe domain
configuration  is very sensitive to thickness
variations.\cite{huber} Lateral changes in sample thickness impose
lateral changes in the period of  magnetic stripe domains
$\lambda$, in the amplitude of the out of plane oscillation
$\Delta M_{out}$ and, also, in the average in-plane magnetization
component $M_{inplane}$ (see sketch in Fig. 1(b)). Therefore, the
continuous film breaks up in a set of linear parallel regions with
different equilibrium magnetic stripe domain configurations
coupled through magnetostatic and exchange interaction to their
neighbors. Then, in the same way as the periodic magnetic stripe
domain pattern in a continuous film is equivalent to a 2D crystal,
the configuration of magnetic stripe domains in a PMA-MLM can be
considered analogous to a strained superlattice made up of
alternating layers of material with different lattice
parameter.\cite{strained,strained2} Thus, "misfit" between
magnetic stripe domains at the different thickness regions becomes
an essential parameter to understand the physics of this system.
It has been shown\cite{aurelio} that the consequences of the
PMA-MLM fabrication process are to create an in-plane shape
anisotropy and, also, to introduce "edges" within the continuous
layer in which topological defects could be nucleated (e.g. 1/2
vortices in the in-plane magnetization or grain boundaries within
the magnetic stripe pattern). A good control of the essential
parameters needed for nucleation of these topological defects
would open the route to understand defect interactions on an
individual basis (in contrast with previous statistical studies in
disordered
patterns\cite{stripes1,asciutto,defeo,huang,saratz,venus}) and,
also, to study the physics of nucleation and propagation of the
observed fractional topological defects (coupled 1/2 disclination
-1/2 vortex) which is interesting for magnetic logic
devices.\cite{parkin,parkin_new} However, in ref.
\onlinecite{aurelio}, the controlled nucleation of topological
defects was only observed under specific geometrical parameters
and magnetic history conditions.

In this work, we have studied magnetization reversal processes of
weak PMA-MLM's as a function of nanostructure geometry in order to
establish the conditions needed for topological defect nucleation
within their magnetic stripe domain patterns. First, we have
performed a detailed characterization of the magnetic domain
configuration of the PMA-MLM's during magnetization reversal for
different in-plane field orientations and different values of the
lateral multilayer period. Then, an analytical model is proposed
that takes into account magnetostatic, exchange and misfit
interactions between the different patterned regions together with
the coupling between in-plane and out-of-plane magnetization
components. Finally, the interplay between these different factors
has been analyzed as a function of PMA-MLM geometrical parameters
in order to determine the most favorable magnetization reversal
regimes for controlled nucleation of grain boundaries and
disclinations within the magnetic stripe domain configuration.

\section{EXPERIMENTAL}
Amorphous 80 nm NdCo$_5$ alloy films have been grown by
co-sputtering from pure Nd and Co targets on 10 nm Al/Si(100)
substrates, and protected from oxidation with a 3 nm Al capping
layer.\cite{cid} At room temperature, the saturation magnetization
is $M_s = 1100$ emu/cm$^3$ and the perpendicular anisotropy
constant is $K_N \approx 10^6$ erg/cm$^3$,\cite{cid} so that $Q =
K_N/2\pi M_s^2\approx 0.18$ implying that the Nd-Co films can be
considered within the weak PMA regime. The continuous films also
present a small in-plane uniaxial anisotropy induced by the
cosputtering process.\cite{fernando_jap} Then, several e-beam
lithography, lift-off and ion beam etching processes have been
performed in order to create the desired thickness modulation over
an extended sample area.\cite{aureliojpd} The result is a set of
$70 \mu$m $\times 70 \mu$m Nd-Co squares with alternate linear
regions of thickness $t_1 = 50$ nm and $t_2 = 80$ nm, width $w/2$
and lateral period $w$. The patterned grooves are parallel to one
of the square sides, as sketched in Fig. 1(b) and, also, to the
growth induced easy axis. Due to the thickness dependence of
stripe period and in-plane magnetization they will take different
values in the different thickness film regions created by the
grooved topography. In the following we will refer to the period
of magnetic stripe domains and in-plane magnetization component in
the thin and thick regions as $\lambda_1$, $M_1$ and $\lambda_2$,
$M_2$, respectively. A series of samples with $w =$ 0.5, 1, 1.4
and 2 $\mu$m has been fabricated on the same substrate, in order
to analyze the different magnetization reversal regimes as a
function of sample geometry. In the following, they will be
referred to as PMA-MLM($w$) with $w$ the lateral period in $\mu$m.
Flat $70 \mu$m Nd-Co squares of thickness $t_1 = 50$ nm and $t_2 =
80$ nm have also been defined near the nanostructured squares for
control purposes.

The magnetic properties of the PMA-MLM's have been characterized
by focused Kerr magnetometry using a NanoMOKE2$^\circledR$ system
in the longitudinal Kerr configuration to obtain the in-plane
hysteresis loops. Stripe domain configuration during magnetization
reversal has been measured by Magnetic Force Microscopy (MFM) with
a Nanotec system that allows us to apply in-plane variable fields
up to 1 kOe.\cite{aurelio} Domain structure for the in-plane
magnetization component has been obtained with a high resolution
Kerr microscope from Evico Magnetics Gmbh in a longitudinal Kerr
effect configuration.

\section{MAGNETIC CHARACTERIZATION OF WEAK PMA MAGNETIC LATERAL MULTILAYERS}

The characterization of stripe domain configuration in PMA-MLM's
has been performed using two different in-plane applied field
orientations: first, with $H$ parallel to the patterned grooves
(easy axis) and, then, with $H$ perpendicular to them (hard axis).
The first one, easy axis magnetization reversal, will allow us to
obtain the basic magnetic behavior of the patterned sample in a
simple geometrical configuration in which in-plane magnetization
reversal occurs mainly by domain wall motion. The second one, hard
axis magnetization reversal, favors rotation processes of the
in-plane magnetization component. This results in more complex
stripe domain configurations that will allow us to control the
nucleation of topological defects within the system.

\begin{figure}[ht]
 \begin{center}
\includegraphics[angle=0,width=0.8\linewidth]{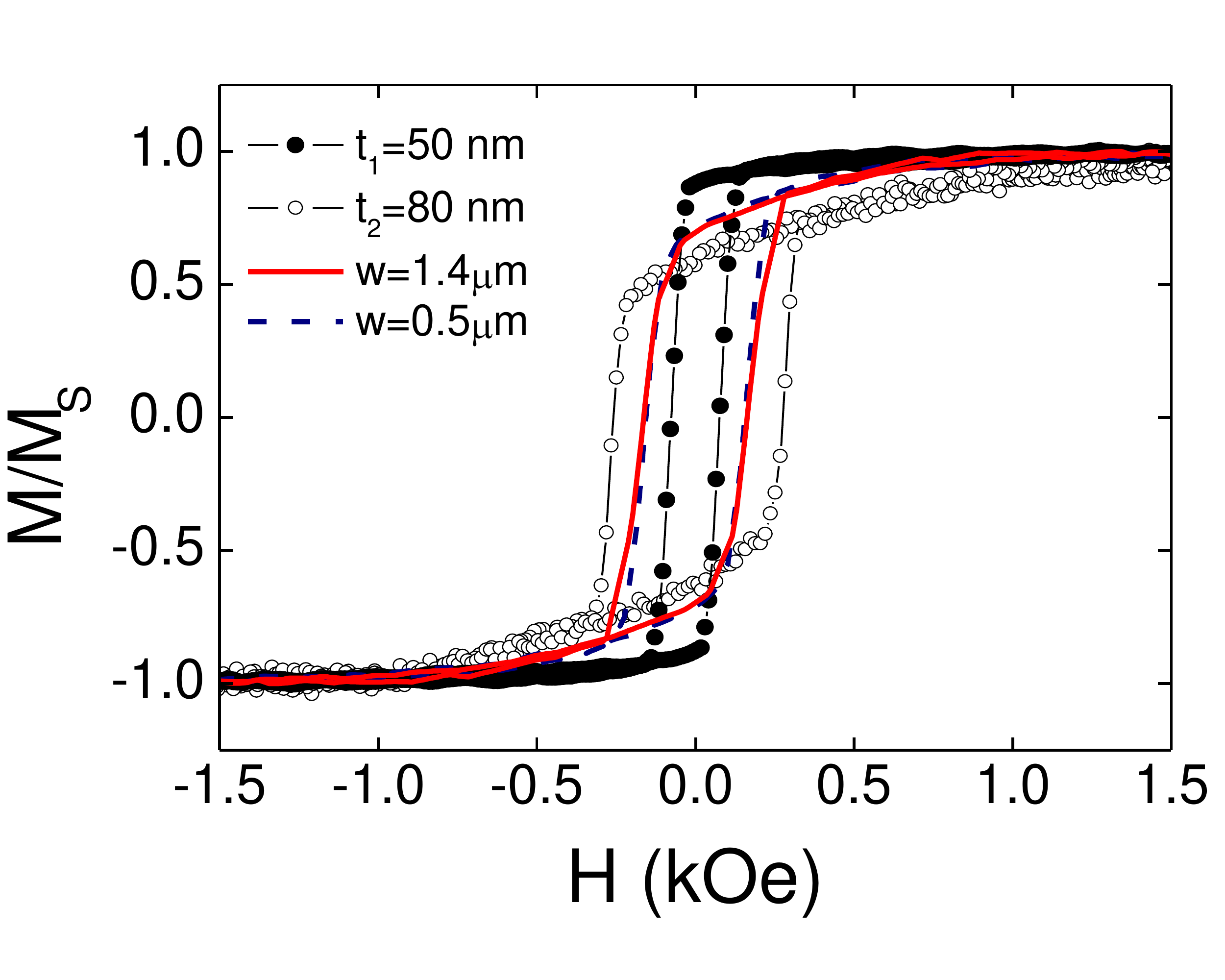}%ahierro_prb_fig2.eps}
 \end{center}
 \caption{(color online) In-plane easy axis MOKE hysteresis loops measured at: $\bullet$, flat square with $t_1 = 50$ nm; $\circ$ flat square with
$t_2 = 80$ nm; solid line, PMA-MLM(1.4); dashed line,
PMA-MLM(0.5).} \label{Fabricacion}
 \end{figure}

\subsection{Easy axis magnetization reversal }

Figure 2 shows the in-plane easy axis hysteresis loops of two
PMA-MLM's in comparison with reference flat squares of thicknesses
50 nm and 80 nm (i.e. equivalent to the thick and thin regions in
the MLM's). All the loops present qualitatively the same
transcritical shape, typical of weak PMA materials, with a reduced
remanent magnetization followed by an almost linear approach to
saturation as the magnetization rotates within the stripe domains
towards the in-plane applied field direction.\cite{huber,loop} The
main differences appear in the remanent magnetization $M_R$ and
coercivity $H_C$ values: the thicker 80 nm flat square shows the
lowest $M_R = 0.6 M_S$ and largest $H_C$ = 260 Oe, whereas the
thinner 50 nm flat square displays the largest $ M_R= 0.88M_S$ and
smallest $H_C=$80 Oe, which is the trend expected from the
thickness dependence of these parameters in weak PMA
films.\cite{fernando_jap} The two MLM's present an intermediate
behavior with $M_R \approx 0.7M_S$ and $H_C \approx$ 160 Oe. This
could be taken, as a first approach, as an indication that the
effect of patterning is equivalent to creating an intermediate
effective thickness in between $t_1$ and $t_2$.

\begin{figure}[ht]
 \begin{center}
 \includegraphics[angle=0,width=1\linewidth]{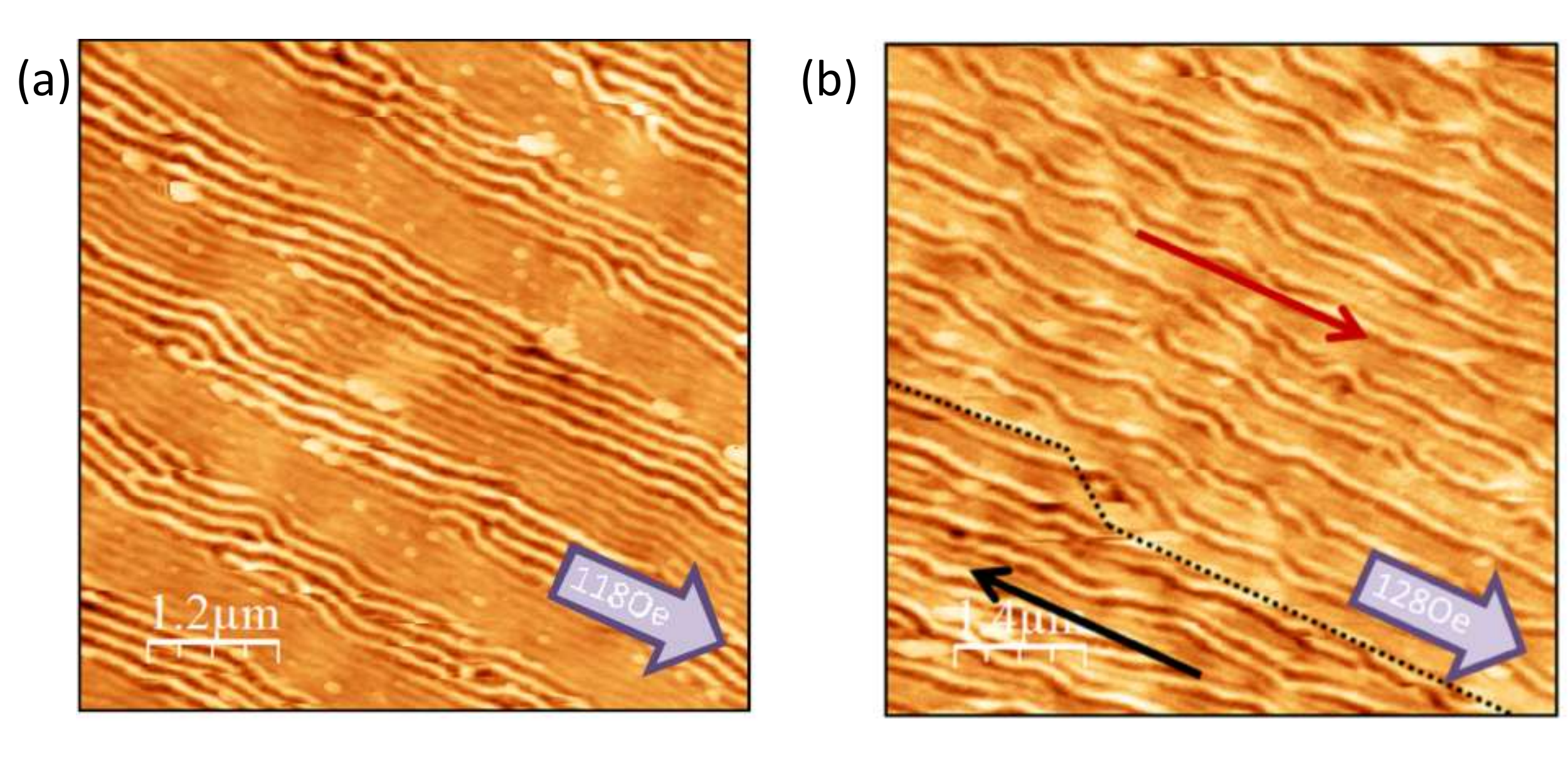}%ahierro_prb_fig4_rev.eps}
 \end{center}
 \caption{(color online) MFM images taken during an in-plane easy axis magnetization reversal process for (a) PMA-MLM($1.4$) at H = 118 Oe
 and (b) PMA-MLM($0.5$) at H = 128 Oe. Thick arrows indicate applied field direction. Thin arrows
 and dashed line indicate in-plane magnetization orientation and
 domain wall position extracted from the analysis of  MFM
 contrast (see Supplemental material\cite{Supplemental}).}
 \end{figure}

However, the detailed MFM characterization of PMA-MLM(1.4) and
PMA-MLM(0.5) reveals a clear influence of the lateral multilayer
structure in stripe domain configuration. Figures 3(a-b) are 6
$\mu$m $\times$ 6$\mu$m MFM images of PMA-MLM(1.4) and
PMA-MLM(0.5), respectively, taken after saturating them with an
in-plane $H = -1$ kOe parallel to the nanostructured lines and,
then, applying a positive field along the same direction, close to
the coercivity. They display the typical stripe domain pattern of
weak PMA films aligned with the last saturating field
orientation.\cite{field} The effect of thickness modulation can be
seen in the different magnetic stripe periods measured at the thin
and thick regions\cite{aurelio} with $\lambda_1 = 130$ nm
 and $\lambda_2 = 160$ nm for PMA-MLM(1.4), while $\lambda_1 \approx 110-120$ nm
 and $\lambda_2 = 170$ nm for PMA-MLM(0.5). It is interesting to note that, in this second case,
 $\lambda$ values are comparable to $w = 0.5 \mu$m, the PMA-MLM period, so that only a couple of stripe domains
 fit within each nanostructured line.

A detailed characterization by Kerr microscopy and MFM reveals
qualitative changes in
 the easy axis magnetization reversal process as the lateral multilayer period is reduced as shown in ref.
 \onlinecite{Supplemental}:  in the
sample with a larger lateral period $w = 1.4 \mu$m, the effect of
patterning is to separate the sample into a set of independent
linear regions that switch by the propagation of head-to-head
domain walls along them whereas in the sample with the smaller
lateral period $w = 0.5 \mu$m, coupling between the different
regions dominates and magnetization reversal is more coherent.

\subsection{Hard axis magnetization reversal }

\begin{figure*}[ht]
 \begin{center}
\includegraphics[angle=0,width=0.65\linewidth]{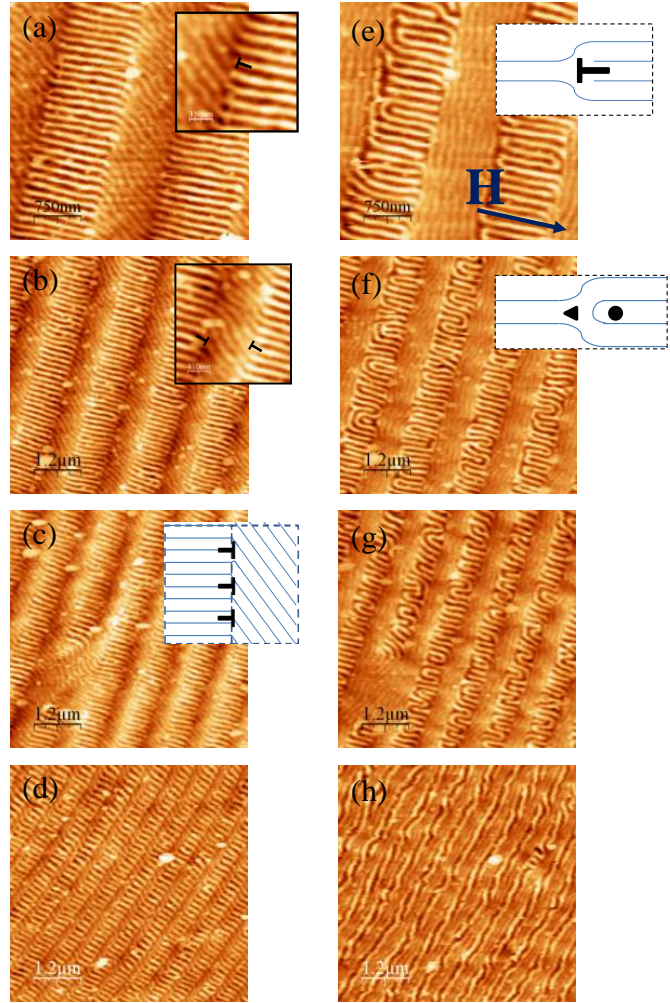}%ahierro_prb_fig5.eps}%MFMsketch_bn.eps}
 \end{center}
 \caption{(color online) MFM images taken during a hard axis magnetization reversal process at remanence: (a) PMA-MLM(2), (b)
 PMA-MLM (1.4), (c) PMA-MLM (1), (d) PMA-MLM (0.5) and close to the coercivity (H=100Oe):(e) PMA-MLM(2), (f)
 PMA-MLM (1.4), (g) PMA-MLM (1), (h) PMA-MLM (0.5). Insets in (a-b) are zooms to
 highlight misfit dislocations. Inset in (c) is a sketch of a low angle boundary
 made up of a equispaced array of misfit dislocations in between two stripe domain patterns with different $\lambda$ and orientation.
 Inset in (e) is a sketch of a dislocation with Burgers vector $|b| = 2\lambda$. Inset in (f) is a sketch of the corresponding dissociated
  disclination pair made up of a "dead end" (+1/2 disclination, $\bullet$) and a "branch" (-1/2 disclination, $\blacktriangle$).}
 \end{figure*}

The behavior of PMA-MLM's is much more complex when a magnetic
field is applied in-plane and perpendicular to the nanostructured
lines, and very interesting confined labyrinthine configurations
appear within the stripe domain pattern due to the competition
between the different anisotropy terms acting on the system.
Figure 4 shows several pairs of MFM images taken along a hard axis
magnetization reversal process in the different PMA-MLM's studied
in this work. The samples have been saturated first with $H=-1$
kOe applied in-plane perpendicular to the patterned lines; then,
the field has been reduced to zero (Figs. 4(a-d)); and, finally,
it has been increased up to $H = 100$ Oe, corresponding
approximately to the hard axis coercivity (Figs. 4(e-h)).

All the samples present a similar remanent stripe domain
configuration (Figs. 4(a-d)): stripes at the thick regions remain
perpendicular to the lines (i.e. along the direction of the last
saturating field, as is the usual case in PMA materials due to
rotatable anisotropy) but stripes at the thin regions have rotated
away from the field direction towards the in-plane easy axis
defined by the shape anisotropy created by the artificial
thickness modulation. It can be seen that a number of "misfit"
dislocations have been generated at the interfaces (see insets in
Figs. 4(a-b)). Due to the differences in equilibrium $\lambda_i$
and stripe orientation in thick and thin regions, stripe periods
projected along the interface are also different (see sketch in
Fig. 4(c)). Thus, there are extra stripes that terminate on the
interface and misfit dislocations are created. Actually, since
this happens on a periodic basis, the array of equispaced misfit
dislocations can be considered analogous to a low angle grain
boundary within the magnetic stripe pattern.

Then, upon applying a reverse perpendicular magnetic field (Figs.
4(e-g)), the rotation process continues within the thin regions
until stripe domains become aligned to the nanostructured lines.
However, in the PMA-MLM's with $w \geq 1 \mu$m, stripe domains
within the thick regions are still perpendicular to the
nanostructured lines so that a set of 90$^0$ boundaries has been
induced within the magnetic stripe domain pattern of these
PMA-MLM. The configuration of these boundaries is quite different
from the low angle boundaries observed in the remanent MFM images.
The 90$^0$ boundaries are decorated by high Burgers vectors
dislocations and dissociated 1/2 disclination pairs. Sketches of
these topological defects are shown in the insets of Fig. 4(e) for
a dislocation with Burgers vector modulus $|b| = 2\lambda$) and of
Fig. 4(f) for a dissociated disclination pair made up of a +1/2
disclination ("dead end") and a -1/2 disclination ("branch") that
is equivalent to a dislocation with $|b| = 2\lambda$. The typical
size of the observed disclination pairs is in the range $\lambda$
to $3\lambda$. These higher energy topological defects are needed
to relieve the large mismatch in between the projected stripe
periods at both sides of the boundary due to their almost
perpendicular orientation.\cite{huang} It is interesting to
mention that it has been shown that +1/2 disclinations in the
magnetic stripe domain pattern are coupled to half vortices in the
the closure domain structure for in-plane magnetization, that
appears along the magnetization reversal process of the thick
lines.\cite{aurelio}

Finally, it must be noted that a different behavior is found in
the stripe domain configuration at coercivity for PMA-MLM(0.5)
(Fig.4(h)): the whole stripe domain pattern in both thin and thick
regions has rotated away from the applied field direction and is
now aligned with the nanostructured lines. Thus, the stripe
configuration becomes much simpler without the high angle
boundaries present in the larger lateral period MLM's.

These different behaviors as a function of lateral multilayer
period can be seen in more detail in Fig. 5 in which the field
dependence of stripe domain orientation relative to the patterned
lines is shown for PMA-MLM(1.4) and PMA-MLM(0.5) during a hard
axis hysteresis loop (squares and circles correspond to stripes in
the thin and thick regions respectively). The angular orientation
data have been extracted from the Fast Fourier Transforms (FFT) of
a series of consecutive MFM images taken during the hard axis
magnetization reversal process. Briefly, topography images
recorded simultaneously with the MFM signal are used as a mask to
divide each image in two, corresponding to the thin and thick
regions. Then, the FFT of each image is used to obtain the angular
orientation of stripe patterns within each kind of patterned lines
in a precise way.\cite{aurelio} For the PMA-MLM with wider lateral
multilayer period (Fig. 5(a)), stripe domains in the thin regions
start close to the negative perpendicular orientation at negative
fields (which corresponds to the saturated state for in-plane
magnetization in this hard axis loop) and, then, perform a
continuous rotation towards the positive perpendicular direction
as the hard axis field intensity
 increases. These rotations within the stripe pattern are directly linked to the rotation that in-plane magnetization
 performs under the applied field torque.\cite{aurelio} It is interesting to note the small overshoot that appears in
 this rotation process up to $\theta \approx 100^0$ before the stripe domain orientation stabilizes close
 to $\theta = 95^0$ for large positive hard axis fields, which is the typical behavior in Stoner-Wolfarth
 rotation processes under an applied field slightly misaligned with the uniaxial anisotropy hard axis (by $5^0$ here).
  At the same time, stripes in the thick regions retain their original perpendicular orientation during the whole measured
   field range (except for a possible $\pm 90^0$ indetermination). On the other hand, for PMA-MLM(0.5) (Fig. 5(b)),
   both thin and thick regions start close to negative saturation (i.e. close to $\theta = -90^0$) at negative fields
    but with a 30$^0$ angular difference. Then, as the positive hard axis field increases, stripes in both kinds of
    regions begin to rotate towards the easy axis, reducing the angular distance between them. Once they
    reach $\theta =0^0$, the stripe domains in the whole sample become
coupled and rotate in unison for the rest
    of the hard axis hysteresis loop until they reach the positive saturation orientation $\theta = 90^0$.

 \begin{figure*}[ht]
 \begin{center}
\includegraphics[angle=0,width=0.8\linewidth]{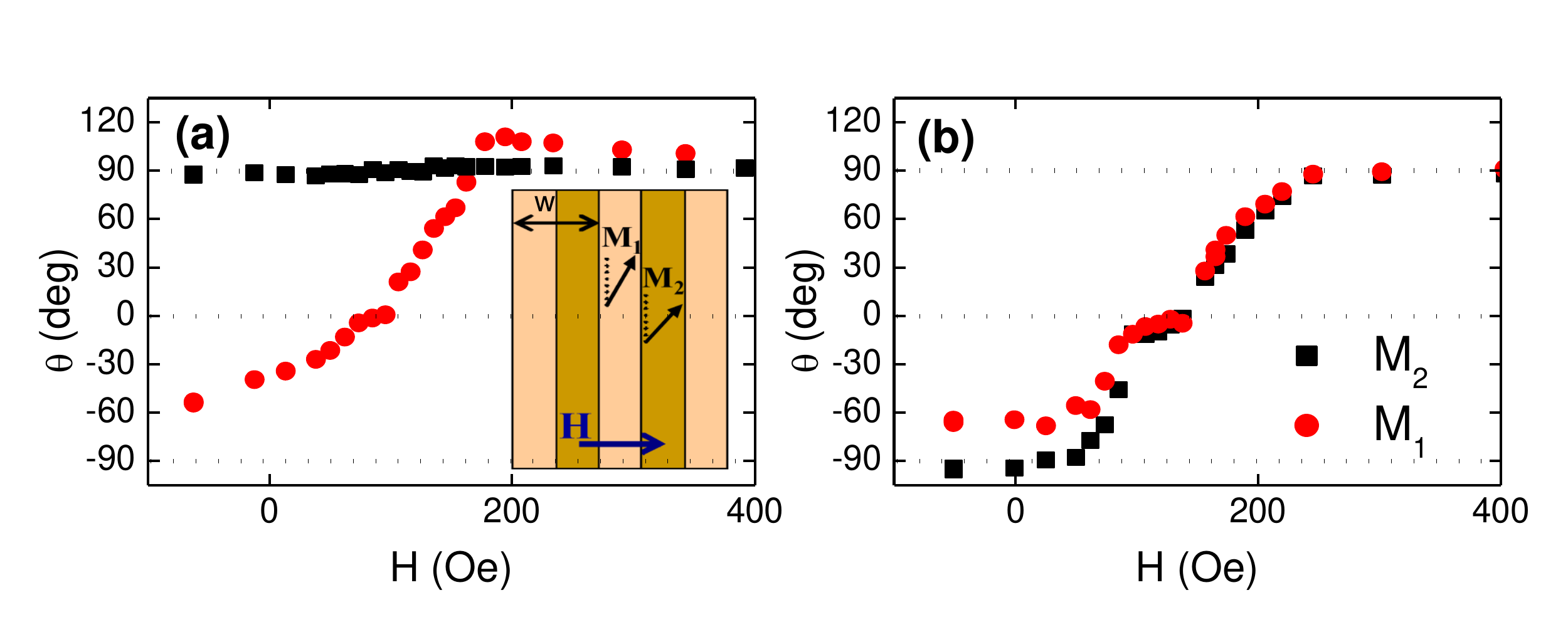}%ahierro_prb_fig6.eps}
 \end{center}
 \caption{(color online) Field dependence of the angular
 orientation of stripe domains (and, consequently, in-plane magnetization)
 during an in-plane hard axis loop measured with the field increasing from negative to positive saturation for (a) PMA-MLM($1.4$) and (b) PMA-MLM($0.5$): $\blacksquare$, thin lines; $\bullet$, thick lines.
 Sketch indicates the angles measured relative to the lines direction.}
 \end{figure*}

We may summarize the results of the MFM characterization in two
points: first, the nucleation of high angle grain boundaries
within the magnetic stripe domain pattern occurs mainly along hard
axis magnetization reversal processes. In this geometry, rotation
of the in-plane magnetization component becomes the preferred
magnetization reversal mode in the thin regions. Due to the
coupling between in-plane magnetization and stripe domains, the
hard axis field acts as a handle to rotate the magnetic stripe
domains in selected areas. Thus, grain boundaries are nucleated
within the magnetic stripe pattern at the limit between different
thickness regions. Second, in the PMA-MLM's with larger lateral
period $w \geq 1 \mu$m, grain boundary angle increases up to
$90^0$ at the hard axis coercivity and 1/2 disclination pairs are
observed. A more coherent behavior appears in PMA-MLM(0.5) both
along the easy axis and hard axis magnetization reversal
processes. This indicates the stronger role of coupling between
neighboring lines as the structural period $w$ is reduced down to
values comparable to the magnetic stripe period $\lambda$.

\section{ANALYTICAL MODEL OF WEAK PMA MAGNETIC LATERAL MULTILAYERS}

\subsection{Analytical model}
The magnetic characterization of PMA-MLM's has shown different
possibilities to control stripe domain configuration making use of
the coupling between in-plane magnetization and stripe patterns in
weak PMA films. Now, in order to understand the observed
experimental conditions for topological defect nucleation and
their dependence on lateral multilayer geometrical parameters, we
must consider the interplay between the different energy terms
involved in the system.

 A first approach to analyze the magnetization rotation in the thin regions during
  a hard axis reversal process in the larger period PMA-MLM's can be made using a
  simplified model for in-plane magnetized MLM's.\cite{aurelio,lateral} In this case,
  only dipolar and Zeeman energy terms related to in-plane magnetization components are considered ($M_1$
  in the thin lines and $M_2$ in the thick lines): the effective shape anisotropy created
   by the flux discontinuities
 that appear at the interface between thin and thick lines due to the lateral
 modulation of the in-plane magnetization,\cite{lateral} and the Zeeman energy of $M_{1}$, assuming that $M_2$ is fixed at $90^0$.
 In this framework, the energy density $e_1$ at the thin lines
 for $M_1$ oriented at $\theta_1$ relative to the lines and $H$ at $\theta_0$ may be written as:
\begin{equation}
e_1 = 2\pi N_x (M_1\sin\theta_1-M_2)^2 - H M_1
\cos(\theta_1-\theta_0)
\end{equation}
with $N_x$ the demagnetizing factor perpendicular to the lines.

However, this simple model is not enough either to capture the
physics of the lateral period dependence of the magnetic behavior
observed in section III nor to give information about the
fabrication parameters needed to create variable angle boundaries
within the stripe domain pattern of PMA-MLM's. A more complete
analytical model should also take into account the energy density
associated to the stripe domain pattern $e_{\bot}$, exchange and
magnetostatic energy terms associated to in-plane magnetization
components $e_{in plane}$, and the coupling between in-plane and
out-of plane magnetization components, that takes the form of a
rotatable anisotropy.

Regarding $e_{\bot}$, it has been shown that the magnetic energy
of the stripe domain pattern can be written as the effective
elastic energy\cite{abanov} of a 2D crystal in terms of the
deformations relative to the equilibrium periodic stripe domain
configuration at a given field. Within this framework, PMA-MLM's
can be considered as the 2D equivalent of strained superlattices
since their stripe domain pattern is composed of alternating
regions with different equilibrium stripe period,
$\lambda_i^{eq}$.\cite{aurelio} Then, we may write their effective
elastic energy as\cite{strained}
\begin{equation}
e_{\bot} = 1/2 B_1 \delta_{el,1}^2+ 1/2 B_2 \delta_{el,2}^2
+2\gamma_{GB}/w
\end{equation}
where $B_i$ is the effective bulk elastic modulus of the stripe
domain pattern within each region; $\delta_{el,i}$ is the elastic
strain in region $i$ due to the difference between the equilibrium
$\lambda_i^{eq}$ and its actual value $\lambda_i$ so that
$\delta_{el,i} = (\lambda_i-\lambda_i^{eq})/\lambda_i^{eq}$;
finally, $\gamma_{GB}$ stands for the energy of the grain
boundaries between stripe domains in thin and thick regions.
$\gamma_{GB}$ depends on the misfit strain $\delta_{misfit}$
between the stripe domain patterns at both sides of the boundary,
\begin{equation}
\delta_{misfit}=(\frac{\lambda_1}{\sin\theta_1}-\frac{\lambda_2}{\sin\theta_2})/\frac{\lambda_2}{\sin\theta_2},
\end{equation}
since $\frac{\lambda_i}{\sin\theta_i}$ is the stripe pattern
period at each region, projected along the grain boundary plane.
For small enough $\delta_{misfit}$, the grain boundary can be
considered as an array of equispaced misfit dislocations with
Burgers vector modulus $|\textbf{b}| = \lambda_2$ located at
$\lambda_2/|\delta_{misfit}|$ distance (see sketch in Fig. 4(c)).
Then,
\begin{equation}
 \gamma_{GB} = |\delta_{misfit}|e_{dis}/\lambda_2,
\end{equation}
 with $e_{dis}$ the energy of a single dislocation given by $e_{dis} = G \lambda_2^2\ln(\alpha w/2\lambda_2)$,
  where $G$ is the dislocation energy coefficient and $\alpha$ is a constant.\cite{strained}

The energy for in-plane energy magnetization components $e_{in plane}$ in a MLM can be written as\cite{lateral}
\begin{equation}
e_{in plane} = 2\pi N_x (M_1\sin\theta_1-M_2\sin\theta_2)^2 +
2\gamma_{DW}/w
\end{equation}
which generalizes eq. (1) to take into account $\gamma_{DW}$, the
energy of the domain wall that appears in between thin and
 thick regions due to the different in-plane magnetization orientations. In general, $\gamma_{DW}$ is a function of $\theta_1-\theta_2$. We have taken,
 as a first approach, $\gamma_{DW} \approx \gamma_0 (1-\cos(\theta_1-\theta_2))$, considering that $M_1$ and $M_2$
 are constant throughout each patterned line so that exchange takes place primarily at the interfaces.\cite{huber}

Finally, coupling between in-plane and out-of-plane magnetization is given by "rotatable magnetic" anisotropy. Briefly, when a
PMA film has been saturated in plane, and the applied field
intensity decreases along a hysteresis loop, stripe domains are
nucleated as a weak out-of-plane oscillation parallel to in-plane
magnetization and, correspondingly to the applied field direction.
As $H$ goes down to zero, the amplitude of this out-of-plane
oscillation increases and the in-plane magnetization component
$M_{in plane}$ is reduced. As a consequence, possible in-plane
magnetization rotations are hindered by the need to reorient the
whole stripe pattern and a "pseudo uniaxial" in-plane anisotropy
is created in the system. The last saturating field direction
becomes an in-plane easy axis with its corresponding anisotropy
constant $K_{rot}$, that can be estimated as\cite{luis}
\begin{equation}
K_{rot} = 4\pi M_S^2 J_2^2(\beta_0)\{1-\frac{\lambda}{4\pi
t}[1-\exp(-\frac{4\pi t}{\lambda})]\},
\end{equation}
where $J_n$ are the Bessel functions of the first kind and
integral order $n$, $\lambda$ is the stripe pattern period, $t$ is
the film thickness and $\beta_0$ is given by the implicit
condition $M_{in plane}/M_S = J_0(\beta_0)$. $K_{rot}$ has a
magnetostatic origin, thus it is proportional $2\pi M_S^2$. It is
zero for $M_{in plane} = M_S$, i.e. $\beta_0=0$ and no stripe
pattern, and increases as $M_{in plane}$ decreases, as shown in
Fig. 6(a).

We will begin our analysis by considering only the interplay
between rotatable anisotropy (that accounts for the coupling
between in-plane and out-of-plane magnetization components in weak
PMA materials) and shape anisotropy (that accounts for the effect
of lateral patterning on in plane magnetization) in section IV.B.
These two terms have a common magnetostatic origin and it will be
shown that their competition captures the essential physics to
understand the nucleation of grain boundaries within the magnetic
stripe domain patterns of the MLM. Then, in section IV.C, we will
turn our attention  to the remaining energy terms related with in
plane domain walls and grain boundaries in the stripe pattern.
This, will allow us to understand the transition to the strongly
coupled regime observed for PMA-MLM(0.5). All these analysis will
be performed considering that the system is at remanence for
simplicity.

\subsection{Nucleation of grain boundaries within the stripe pattern: Competition between shape and rotatable anisotropies}

As a starting point to study the competition between rotatable
anisotropy and shape anisotropy it is interesting to consider the
behavior of a single weak PMA infinite nanowire of thickness $t$
and width $w/2$. In this simplified case, $e_{in plane}$ reduces
only to shape anisotropy energy and eq. (5) becomes
\begin{equation} e_{inplane} = 2\pi N_x M_{in
plane}^2\sin^2\theta =  K_{shape}\sin^2\theta
\end{equation}
 where $K_{shape}$ is the shape anisotropy constant and $2\pi N_x = 4 \arctan (4t/w)
- (w/2t)\ln(1+16t^2/w^2)$.\cite{Nx}  $K_{shape}$ increases as a
function of $M_{in plane}$ and is enhanced as nanowire
 width decreases, as shown in Fig. 6(a).

 \begin{figure}[ht]
 \begin{center}
\includegraphics[angle=0,width=1\linewidth]{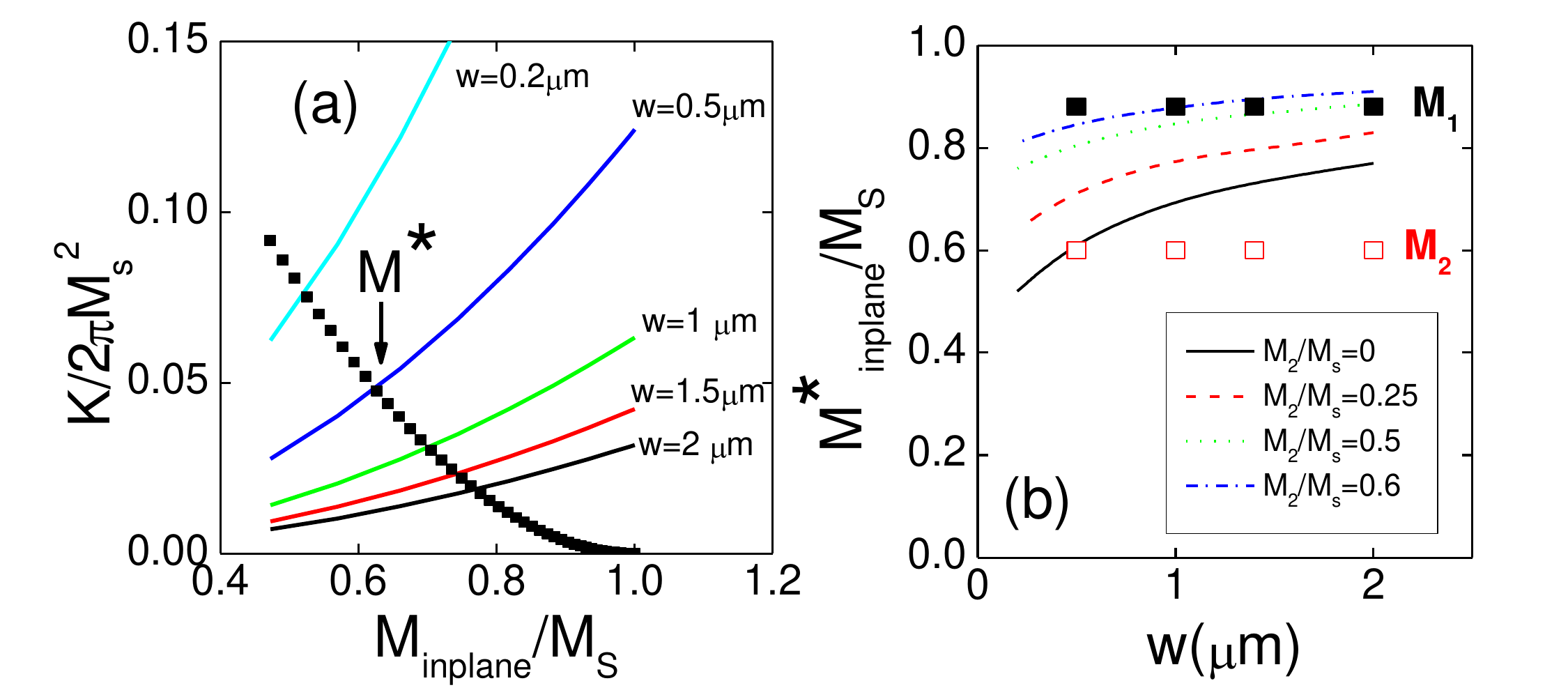}%ahierro_prb_fig7.eps}
 \end{center}
 \caption{(color online) (a) Dependence of magnetic anisotropy with in-plane magnetization component for a PMA nanowire of thickness $t_1$ = 50 nm and width w/2: squares, $K_{rot}$; solid lines, $K_{shape}$.
 (b) Dependence of $M^{*}_{in plane}$ on LML period for different values of the magnetization in the thick lines.
 Symbols indicate $M_1$ and $M_2$ remanent values for the thin and thick
 regions extracted from the hysteresis loops of the continuous control films. }
 \end{figure}

 The effect of shape anisotropy on the system is to
rotate $M_{inplane}$ towards the wire axis. However, for a wire
magnetized perpendicular to the wire axis,
 $K_{shape}$ would compete with the rotatable anisotropy $K_{rot}$ that tends to keep the stripe pattern (and its associated $M_{in plane}$) in its original orientation:
 for small $M_{in plane}$ and large $w$, $K_{rot}$
should dominate the
 nanowire magnetic behavior and no rotations should appear during a
 hard axis hysteresis loop. On the contrary, a regime dominated by $K_{shape}$ would appear for large enough $M_{in plane}$
 and/or small nanowire width in which stripe domains rotate during hard axis magnetization reversal following $M_{in
 plane}$ until they
 become aligned with the nanowire axis at coercivity. The
 crossover between these two regimes can be calculated, as a first
 approximation, by the condition $K_{rot}=2\pi N_x M_{in plane}^2$. This defines
 a boundary $M^{*}_{in plane}(w)$ within the $M_{in plane}$ vs. $w$ plane, as indicated by
 the solid line in Fig. 6(b). Stripe domain rotation will be favored in
 nanowires with $M_{in plane}$ and $w$ above this line, whereas stripe domains
 should remain fixed by $K_{rot}$ for wires with parameters below
 it.

 This single wire diagram can be used as a starting point to model the behavior of PMA-MLM's as
 two sets of infinite parallel
 nanowires of width $w/2$, alternating thicknesses $t_1$ and $t_2$ and in-plane magnetizations $M_1$ and $M_2$.
 Magnetostatic coupling between the two sets of nanowires
  due to  the flux
 discontinuities that appear at the interfaces modifies eq.(7), giving
 \begin{equation}
e_{inplane} = 2\pi N_x (M_1\sin\theta_1-M_2\sin\theta_2)^2.
\end{equation}
We have calculated again the crossover condition as $K_{rot}=2\pi
N_x (M_{1}-M_2)^2$, considering that the competition between both
terms occurs when the system is at the initial hard axis saturated
configuration with $M_1$ and $M_2$
 perpendicular to the wires axis. Then, the effect of magnetostatic coupling is to move the crossover lines $M^{*}_{in plane}(w)$
 for the PMA-MLM's to higher $M_{in plane}$ values, as shown in Fig. 6(b), since it reduces
 $e_{shape}$.

Now, we may use this $M_{in plane}$ vs. $w$ diagrams to predict
the
 nucleation of grain boundaries within the stripe pattern of PMA-MLM's:
  for a given value of $w$, $M_1$ and $M_2$ should lie at
different sides of the crossover line between $K_{rot}$ and
$K_{shape}$ dominated regimes so that stripes in thin regions
rotate while stripes in the thick regions remain fixed.
%Thus, in a PMA-MLM's series with fixed $M_1$ and $M_2$ ratio and varying lateral period $w$ as studied here, two different magnetization reversal regimes are predicted:  a "continuous film" limit at large $w$ in which $K_{rot}$ dominates for all thicknesses and no stripe rotations are seen and, below a certain $w^*$, a "grain boundary" limit in which $M_1$ is above the
%$M^{cross}_{in plane}(w)$ line corresponding to $M_2$.
We can take, as an example, the remanent magnetization values of
the control films with $t_1 = 50$ nm  and $t_2 = 80$ nm, indicated
by filled and hollow squares in Fig.6(b) and compare them with the
observed behavior in the PMA-MLM's of Fig.4. $M_2=0.6M_S$ is well
within the "non-rotating" region, whereas $M_1 = 0.88 M_S$ lies
close to the crossover line calculated for $M_2 = 0.6M_S$ changing
from one regime to the other at $w=1$ $\mu$m. This is
qualitatively in agreement with the observation of "grain
boundaries" in the PMA-MLM's with $w$ in the $\mu$m range.
However, it underestimates the maximum lateral period compatible
with "grain boundary nucleation", probably because the start of
the stripe domain rotation process occurs well before remanence
(i.e at larger $M_1$ values).

\subsection{Coupling effects in PMA-MLM's}
One of the limitations of the analysis made in the previous
subsection is that it does not predict the strongly coupled regime
observed in PMA-MLM(0.5), in which the stripe pattern in the whole
sample rotates in unison. Thus, other sources of coupling in
between the different patterned regions must be considered in
addition to magnetostatic coupling.

There are three energy terms in eqs. (2) and (5) that scale as
$1/w$ and should dominate the behavior of the system in the small
$w$ limit: grain boundary energy within the magnetic stripe
pattern $2\gamma_{GB}/w$, domain wall energy within the in-plane
magnetization configuration $2\gamma_{DW}/w$ and magnetostatic
coupling $2\pi N_x (M_1\sin\theta_1-M_2\sin\theta_2)^2$ through
the width dependence of $N_x$. These different terms have been
calculated as a function of lateral multilayer period, as shown in
Fig. 7, using the experimental $\theta_1$ and $\theta_2$ values
obtained from the MFM images taken at remanence during a hard axis
loop (Figs.4(a)-(d)). We have chosen to study the evolution of the
system in the remanent state because of two reasons: first, Zeeman
energy terms need not to be considered, which simplifies the
analysis; second, the MFM images show that at this state, grain
boundaries in between thin and thick regions are composed of a
simple misfit dislocation array so that eq. (4) can be used to
estimate $\gamma_{GB}$.
\begin{figure}[ht]
 \begin{center}
\includegraphics[angle=0,width=1\linewidth]{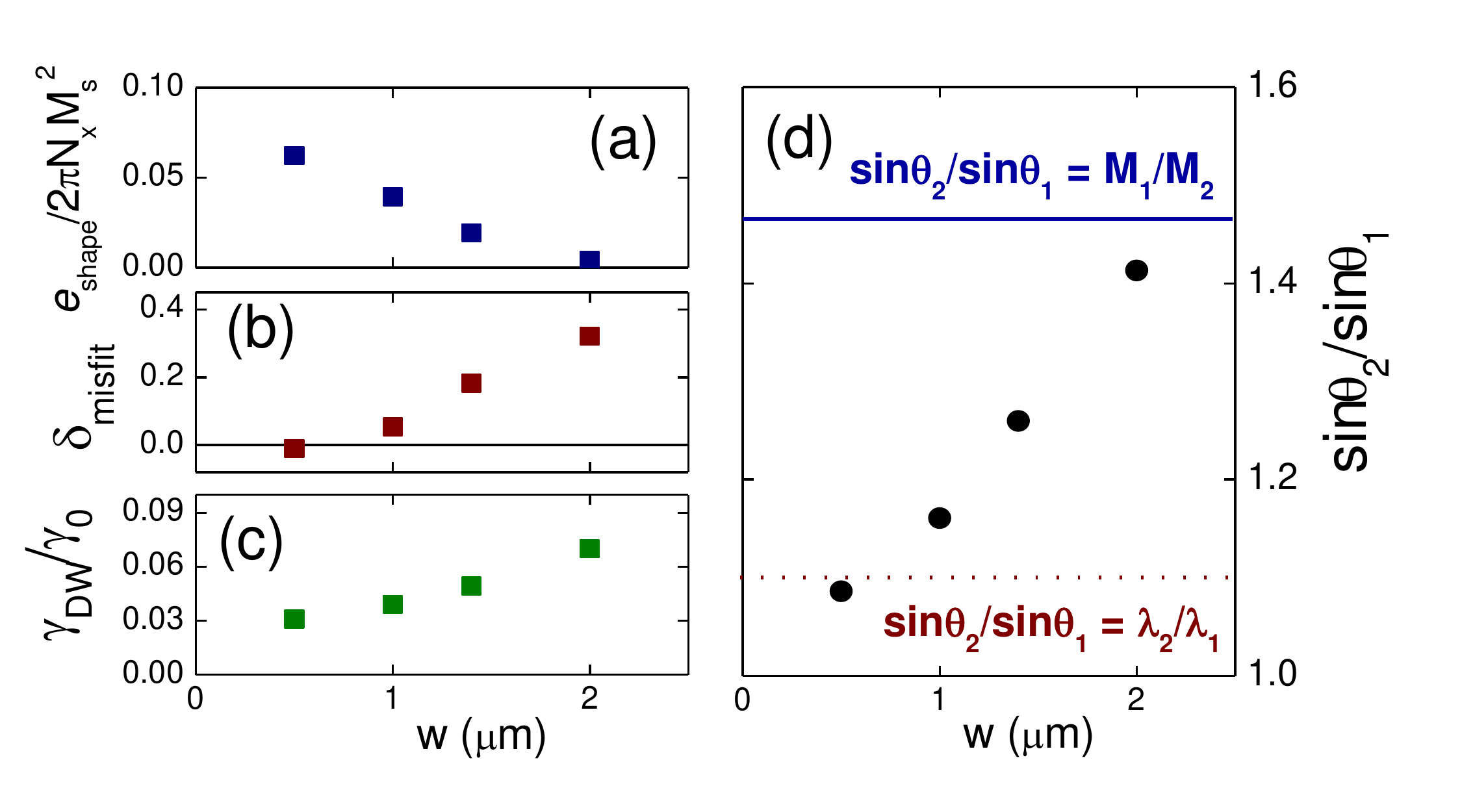}%ahierro_prb_fig8.eps}
 \end{center}
 \caption{(color online) Dependence on LML period of (a) magnetostatic energy term normalized to $2\pi N_x M_S^2$, (b) misfit strain and (c)
 in-plane domain wall energy, calculated from MFM images at remanence. (d) $\sin\theta_2/\sin\theta_1$  vs. $w$.
 Solid and dashed lines correspond to zero magnetostatic energy and zero misfit strain, respectively.}
 \end{figure}

The first thing that can be noticed in Fig.7(a) is that, starting
from PMA-MLM(2), the magnetostatic energy term, normalized by
$2\pi N_x M_S^2$, increases
 as $w$ is reduced. On the other hand, both $|\delta_{misfit}|$, which is proportional to $\gamma_{GB}$, and $\gamma_{DW}$
 decrease as the small $w$ region is approached (Figs. 7(b-c)). This implies that the system prefers to minimize
 these last two interaction terms as $w$ is reduced at the expense of adopting a less favorable  configuration for
 the magnetostatic energy term.

 This can be seen in more detail in Fig. 7(d) in which we have
 plotted the ratio $\sin\theta_2/\sin\theta_1$ vs. $w$ in order to compare it with the conditions that minimize each of
 these interaction terms. First, the magnetostatic energy term will be zero
 if $M_1\sin\theta_1-M_2\sin\theta_2=0$, i.e.
  $\sin\theta_2/\sin\theta_1 = M_1/M_2 = 0.88/0.6 = 1.47$, which is calculated using the remanent magnetization values
  of the control films with thickness $t_1$ and $t_2$. Second, grain boundary energy should be minimum
  if
  $\delta_{misfit} = 0$, which corresponds to $\sin\theta_2/\sin\theta_1 = \lambda_2/\lambda_1 = 1.1$. Finally,
  exchange energy will be minimized if $\theta_2=\theta_1$ so that $\sin\theta_2/\sin\theta_1 = 1$. It is seen that
  whereas PMA-MLM(2) is close to fulfilling the magnetostatic coupling minimum conditions, the ratio
  $\sin\theta_2/\sin\theta_1$ departs towards lower values as soon as $w$ is reduced. Then, when the strongly coupled regime
  of PMA-MLM(0.5) is reached, the condition to minimize $\delta_{misfit}$ is
  fulfilled. This is probably a size effect
since patterned line width $w/2 = 0.25 \mu$m is comparable to
stripe domain period $\lambda \approx 0.1-0.2 \mu$m. Thus, misfit
dislocations cannot be accommodated within the PMA-MLM and stripe
domains at the thin and thick regions become locked in a zero
misfit configuration.

\subsection{Geometrical regimes for stripe domain configuration in PMA-MLM's}
  The results of the analysis of the different energy terms
  involved in these PMA-MLM's show three different
  regimes as a function of lateral multilayer period.

  First, there
  is a large $w$ regime, in which rotatable anisotropy (which is a bulk energy term) dominates and
  the PMA-MLM's behave as continuous unpatterned films with their
  stripe domain patterns oriented along the last saturating field
  direction independently of its orientation relative. This regime would be favored
  both for large $w$ and small difference between $M_1$ and $M_2$, i.e. for small thickness
  modulations, which would explain the absence of rotations found within the stripe patterns of PMA-MLM's with $t_1-t_2 = 12$ nm
  and 0.5 $\mu$m $\leq w \leq 2 \mu$m in
  ref. \onlinecite{aurelio}.

  Second, there is an intermediate $w$ regime,
  in which shape anisotropy induced by the thickness modulation overcomes rotatable anisotropy only in the thin regions
  and, thus, thin and thick regions switch independently during a
  hysteresis process. This is the most interesting regime to study topological defects in the
  magnetic stripe domains, since variable angle grain boundaries and disclinations are nucleated
  at the interfaces, coupled to domain walls for in-plane
  magnetization.

  Finally, there is a
small $w$ regime, in which coupling  between thin and thick
regions becomes strong enough to overcome rotatable anisotropy in
the thick patterned lines and the film switches as a whole during
the magnetization reversal process, under the effect of the
in-plane uniaxial anisotropy induced by the lateral thickness
modulation. In this regime, misfit strain within the magnetic
stripe domain pattern is minimized. This is related with the
similar size of patterned line width $w/2 = 0.25 \mu$m and stripe
domain period $\lambda \approx 0.1-0.2 \mu$m. Thus, misfit
dislocations and disclination pairs needed to nucleate variable
angle grain boundaries in the stripe domain pattern become too
large to fit within the interfaces between thin and thick regions
since their size is of the order $\lambda$ - $3\lambda$, as
observed in the MFM characterization.

In summary, the previous analysis has shown that the essential
physical ingredients needed for controlled nucleation of
topological defects within the magnetic stripe domain
configuration are: first, the existence of localized "misfit
strains" in the stripe pattern created by the local changes of
$\lambda$ due to the nanofabricated thickness modulation; second,
the competition between rotatable anisotropy and shape anisotropy
induced by nanopatterning that allows local rotations of stripe
domains due to their coupling with $M_{in plane}$ and, third, the
use of large enough lateral geometrical features in comparison
with the relevant topological defects to avoid size effects that
would inhibit defect nucleation.

\section{CONCLUSIONS}

In summary, the different regimes of magnetization reversal in
weak PMA-MLM's have been studied as a function of lateral
multilayer period $w$, both experimentally and with the aid of an
analytical model, in order to establish the conditions for
controlled topological defect nucleation within their magnetic
stripe patterns. At $w \geq 1 \mu$m, lateral patterning induces
different reversal processes in the thin and thick regions so that
they switch independently during easy axis and hard axis
hysteresis loops: domain walls are created within the in-plane
magnetization configuration coupled to variable angle grain
boundaries and disclinations within the stripe domain patterns.
This process is driven by the interplay between shape anisotropy
induced by the periodic thickness modulation and the different
values of rotatable anisotropy in the thin and thick regions. On
the other hand, as the lateral period is reduced down to $w = 0.5
\mu$m a strongly coupled regime is found in which the PMA-MLM
switches as a whole and misfit strain within the magnetic stripe
pattern is minimized.

\begin{acknowledgments}
Work supported by Spanish MICINN under grant FIS2008-06249. R. M.
and N. S. acknowledge support from UPV/EHU UFI11/23 and Basque
Country Government grant Etorek SE11-304. A.H-R. acknowledges
support from FCT of Portugal grant (SFRH/BPD/90471/2012).
\end{acknowledgments}

\end{document}